\documentstyle[12pt,epsf,colordvi,aaspp4]{article}

\begin{document}

\title{CROSS-CORRELATING THE SLOAN DIGITAL SKY SURVEY WITH THE MICROWAVE SKY}

\author{Hiranya V. Peiris\altaffilmark{1} and David N. Spergel\altaffilmark{2}}
\affil{Department of Astrophysical Sciences, Princeton University, 
Princeton, NJ 08544}

\altaffiltext{1}{hiranya@astro.princeton.edu}
\altaffiltext{2}{dns@astro.princeton.edu}

\begin{abstract}
While most of the microwave background fluctuations on angular scales greater than a few arcminutes were generated at $z > 800$, the low redshift universe does distort the microwave background.  Since the Sloan Digital Sky Survey (SDSS) traces the structures in the low-redshift universe,  we can gain additional insights into the physics of the low-redshift universe by cross-correlating microwave background maps with template maps produced from the SDSS. We present a formalism for cross-correlating data from the Microwave Anisotropy Probe (MAP) with the Sloan Survey for the thermal Sunyaev-Zel'dovich (SZ) effect, the Integrated Sachs-Wolfe (ISW) effect, and weak lensing. This formalism is used to compute the signal-to-noise for cross-correlating these effects with various classes of tracer objects from the SDSS. It is found that the anticipated samples of SDSS quasars and galaxies with photometrically-determined redshifts would be good tracers for cross-correlating with the CMB. We find that the SZ-galaxy cross-correlation would give good constraints on pressure fluctuations in supercluster-scale gas. Cross-correlating weakly-lensed quasars with maps of the convergence of the CMB is found to give strong constraints on $\Omega_0$ as well as the equation of state $w$. We find that the ISW cross-correlation gives poor signal-to-noise using these techniques.   
\end{abstract}

\keywords{Cosmic microwave background --- galaxies: distances and redshifts
 --- large scale structure of the universe --- gravitational lensing ---
 methods: numerical}


\section{INTRODUCTION} \label{intro}

Microwave background observations and measurements of large scale structure are complementary.  While  the Sloan Digital Sky Survey (SDSS; \cite{gun95}) measures the distribution of galaxies in the nearby universe, NASA's Microwave Anisotropy Probe (MAP; \cite{ben97}) primarily probes the distribution of matter and radiation at z=1300. Since the nearby universe is thought to have the same statistical properties as the distant universe, we can gain insights into low-redshift physics as well as constrain cosmological parameters by cross-correlating the final conditions measured by SDSS with the initial conditions probed by MAP (\cite{bou98}; \cite{eis98}; \cite{wan99}).

To maximize the signal-to-noise from the cross-correlation within the limitations on the CMB data set by the specifications of the MAP Project, a good tracer of the low-redshift density distribution must fulfill two criteria:

\begin{enumerate}
\item{The tracer objects must be numerous enough that the Poisson error term in the cross-correlation is small.}
\item{The tracer must probe as large a volume of redshift-space as possible so that the cross-correlation signal is maximized.}
\end{enumerate}

In practice, one finds that object types with high number densities (e.g. galaxies) tend to have redshift distributions that only probe a small part of the volume and object types that probe a larger fraction of the volume (e.g. quasars) are less numerous. Therefore it is necessary to strike a balance between these two criteria in order to optimize the signal-to-noise. 

In this paper, we investigate several classes of objects that will be available in the SDSS survey in order to find tracers that optimize the cross-correlation with the CMB. We find that SDSS quasars with magnitude $i'<21$ and galaxies with magnitude $r'<21$, which will have photometrically-determined redshifts, are good tracers. We then use the cross-correlation formalism to show that,

\begin{enumerate}
\item{cross-correlating the SDSS galaxy sample with $r'<21$ in various redshift ``slices'' with maps of the Sunyaev-Zel'dovich effect from MAP gives good constraints on the pressure fluctuations in supercluster-scale gas, and}
\item{cross correlating the SDSS quasar sample with $i'<21$ with convergence maps from MAP gives strong constraints on the density parameter and the equation of state.}
\end{enumerate}

The second-order effects that alter the microwave background and the low-redshift tracers used in the cross-correlation are introduced in \S~\ref{zcmb}. In \S~\ref{cc-form} we outline the cross-correlation formalism, derive an optimal filter for maximizing the signal-to-noise from it and discuss briefly the computational issues involved in the cross-correlation. The results of the calculations are presented in \S~\ref{cc-res}. We explore the possibility of constraining the equation of state with these methods in \S~\ref{quin}. Finally, our conclusions are summarized in \S~\ref{finish}.    


\section{THE LOW-REDSHIFT UNIVERSE AND THE MICROWAVE BACKGROUND} \label{zcmb}

There are three primary low redshift effects that alter the cosmic microwave background (CMB).

\subsection{Sunyaev-Zel'dovich (SZ) Effect} \label{sz-desc}

Hot gas in clusters and filaments produce new distortions in the microwave background. The SZ effect (\cite{sun72}) is the change in energy experienced by CMB photons when they scatter from the hot gas. It takes two forms. The dominant contribution is the thermal SZ effect, the gain in energy acquired from the thermal motion of the gas. The kinetic SZ effect arises from the Doppler shift due to the bulk motion of the gas.

\subsection{Integrated Sachs-Wolfe (ISW) Effect} \label{isw-desc}

In low-$\Omega_0$ universes, the fact that potentials decay (in other words, matter fluctuations stop growing) at some epoch makes another contribution to the large-scale CMB anisotropy. In addition to the redshift experienced while climbing out of potential wells on the last scattering surface, photons experience a cumulative change in energy due to the decaying potentials as they travel to the observer. The blueshift of a photon falling into a decaying potential well is not entirely cancelled by the redshift as it climbs out. This leads to a net energy change, which accumulates along the photon path. This Integrated Sachs-Wolfe effect is distinct from the more commonly considered redshifting at the last scattering surface which has become known as the Sachs-Wolfe effect. Both effects are considered in the original paper by \cite{sac67}.

The ISW effect operates most strongly on scales where the change of potential is large over a wavelength. For $\Lambda$CDM models, the effect is confined to the largest angular scales (\cite{kof85}), i.e. $\ell \leq 10$'s.

\subsection{Weak Lensing} \label{lens-desc}

Light propagating through an inhomogeneous universe is distorted through lensing by foreground matter (\cite{gun67}). Dark matter distributed along the line of sight between the surface of last scattering and the present epoch deflects CMB photons and imprints a characteristic pattern onto the CMB anisotropies created by acoustic oscillations. On large and intermediate scales lensing smoothes the acoustic oscillations while on very small scales it creates additional power (\cite{sel96}). 

\cite{zal99} have developed a useful formalism for computing the convergence of the microwave sky. The convergence is a quantity which induces random deflections in the direction of the CMB photons. It is the integrated mass density along the line of sight divided by the critical density.

\subsection{Low-redshift Tracers} \label{tracer-desc}

The SDSS, which will cover about a quarter of the sky, will contain many classes of objects which trace large scale structure. These can be cross-correlated with microwave background maps produced by MAP. Some of these are:

\begin{enumerate}
\item{Photometric-z galaxies ($r' < 21$)}
\item{All galaxies ($r' < 23$)}
\item{Spectrometric-z quasars ($i' < 19$)}
\item{Photometric-z quasars ($i' < 21$)}
\item{FIRST objects}
\end{enumerate}

Here, the magnitude cuts are made in the SDSS $r'$ and $i'$ photometric bands. The first and fourth categories will have photometrically determined redshifts. The magnitude cuts for these categories reflect conservative estimates of the limiting magnitudes to which photo-z techniques can be applied (D. Eisenstein 2000, private communication, G. Richards 2000, private communication). Upper estimates for the expected errors in photometrically-determined redshifts for $r'<21$ galaxies are about 0.04-0.08. Typically, at present, photometric methods for quasars give 50\% of redshifts correct to within $\pm 0.1$ and 70-75\% correct to within $\pm 0.2$. This is fairly independent of magnitude down to $i'=20$. After that, photo-z cannot be done reliably on $z<2.2$ QSOs. However, for $z>2.2$ QSOs, the redshift errors should be comparable. These errors should improve in the near future as details of the photo-z techniques are refined. 

The second category contains galaxies brighter than the $5\sigma$ detection limit of the SDSS r' filter. Since we will not have redshifts for all of these galaxies, the results of cross-correlating this sample with low redshift effects are of academic interest only. The third category contains quasars for which redshifts would be obtained by the SDSS spectroscopic survey. For the SDSS spectroscopic quasar sample, the redshift errors are negligible ($< 0.1\%$). The last category contains objects from the FIRST survey (\cite{whi97}) which would be optically matched by the SDSS.

Approximating the radial distribution function of the tracer sample by a $\delta$-function at the mean redshift of the sample changes cross-correlation results by 5-10\%. We estimate the signal-to-noise to change by roughly the same factor due to photometric redshift errors.


\section{CROSS-CORRELATION FORMALISM} \label{cc-form}

In this section we introduce the notation and develop the formalism for cross-correlating each of the tracer object types with the secondary CMB anisotropies described above.

We begin with the comoving Friedman equation for a spatially flat universe with a cosmological constant $\Lambda$:
\begin{eqnarray} \label{friedman}
\left( \frac{a'}{a} \right) ^2 &=& \frac{8\pi G}{3} \frac{a^2}{c^2} \left[ \frac{\rho_0}{a^3} + \rho_\Lambda \right] \nonumber \\
&=& \frac{a^2 H_0^2}{c^2} \left[\frac{\Omega_0}{a^3} + \left(1- \Omega_0 \right) \right]
\end{eqnarray}
where $a$ is the dimensionless scale factor, prime denotes derivatives with respect to conformal time $\tau$, $H_0$ is the Hubble constant and $G$ is the gravitational constant. We take $\Omega_0 + \Omega_\Lambda = 1$, where $\Omega_0 = 8\pi G \rho_0/(3 H_0^2)$ and $\Omega_\Lambda = 8\pi G \rho_\Lambda/(3 H_0^2)$, where $\rho_0$ and $\rho_\Lambda$ are the mean densities in dark matter and vacuum energy today, respectively. We normalize the scale factor so that $a(\tau_0) = 1$, where $\tau_0$ is the present conformal time. The linear growth factor (\cite{pea99})
\begin{equation} \label{gf}
D\left(\tau\right) \propto \frac{a'}{a^2} \int _0 ^{a} da \left( \frac {a}{a'} \right) ^3 
\end{equation}
is normalized so that $D(\tau_0) = 1$. We now introduce the conformal lookback time $\eta = (\tau_0 - \tau)$, since it is more natural to frame the integrals that follow in terms $\eta$. We have used the conformal transformation $d\tau = (c/a) dt$, where $t$ is the standard time coordinate, with the result that the conformal lookback ``time'' represents the proper motion distance.

The fluctuations in projected surface density of tracer objects $i$ in a redshift slice in direction ${\hat {\bf n}}$ (\cite{moe98}) are given by,
\begin{equation} \label{counts-x}
\Sigma _i ({\hat {\bf n}}) = \int _0 ^{\tau _0} d \eta \left[ b_i \delta ({\bf x}) W(\eta) + 2(2.5 s_i -1) \kappa \right]
\end{equation}
In this expression, the first term in the integrand gives the fluctuations on the sky resulting from intrinsic clustering of the tracer objects. The second term gives the fluctuations arising due to magnification bias in the weak lensing limit, $\kappa \ll 1$: lensing increases the area of a given patch on the sky, thus diluting the number density. On the other hand, tracer objects too faint to be included in a sample of a given limiting magnitude are brightened as a result of lensing and may therefore be included in the sample. The net magnification bias can lead to either enhancement or suppression of the observed number counts, depending on the slope of the number-magnitude relation. 

Here, we have assumed that the tracer density perturbations are biased linearly with respect to dark matter density perturbations, i.e. $\delta_i ({\bf x}) = b_i \delta ({\bf x})$, where for simplicity, the bias factor $b_i$ is assumed constant and independent of scale and redshift. In reality, the bias parameter is likely to be time- and scale-dependent, and could be estimated for real galaxy/quasar samples using their power spectra and factored into the cross-correlation.

W($\eta$) denotes the radial distribution of the tracer objects. $s_i$ is the logarithmic slope of the number counts of tracers $N_0(m)$ in a sample with limiting magnitude $m$,
\begin{equation} \label{slope}
s_i = \frac{d \log N_0(m)}{dm}
\end{equation}
$\kappa$ is the convergence, a weighted projection of the density field along the line of sight:
\begin{equation} \label{kappa}
\kappa = \frac{3}{2} \Omega_0 \left(\frac{H_0}{c}\right)^2 \int_0^{\tau_0} d\eta \frac{g(\eta)}{a(\tau_0-\eta)} \delta ({\bf x})
\end{equation}
where the radial window over the dark matter fluctuations $\delta$ is $g/a$. In a flat $\Omega=1$ universe, the radial weight function is given by,
\begin{equation} \label{radweight}
g(\eta) = \eta \int_{\eta}^{\tau_0} d\eta' \frac{(\eta'-\eta)}{\eta'} W(\eta')
\end{equation}
where $W(\eta')$ denotes the normalized radial distribution of the lensing objects, i.e. the probability of finding an object at a given distance.

Expanding the comoving matter density fluctuations as,
\begin{eqnarray} \label{delta-expand}
\delta ({\bf x}) &=& \delta ({\hat {\bf n}} \eta) \nonumber \\
&=& \int \frac{d ^3 {\bf k}}{(2 \pi)^{3/2}} D(\tau_0 - \eta) \delta ({\bf k}) e^{i {\bf k} \cdot {\hat {\bf n}} \eta}
\end{eqnarray}
Eq.~\ref{counts-x} becomes
\begin{equation} \label{counts-k}
\Sigma _i ({\hat {\bf n}}) = \int _0 ^{\tau _0} d \eta \left[ b_i W(\eta) + 3 \Omega _0 \left( \frac{H_0}{c} \right)^2(2.5s_i -1) \frac{g(\eta)}{a(\tau_0 - \eta)} \right] \int \frac{d ^3 {\bf k}}{(2 \pi)^{3/2}} D(\tau_0 - \eta) \delta ({\bf k}) e^{i {\bf k} \cdot {\hat {\bf n}} \eta}
\end{equation}
We will cross-correlate projected surface density fluctuations of the quasar and galaxy samples from the SDSS, described in \S~\ref{zcmb}, with the CMB.

In the case of the optically matched FIRST sample, we will cross-correlate instead the ellipticity $\epsilon = 2(1-\sigma_\epsilon^2) \kappa$ where the RMS ellipticity error is taken to be $\sigma_\epsilon = 0.4$ (\cite{ref98}). The weak lensing shear $\gamma_i$ is related to the source-averaged 
ellipticity by $\langle \epsilon_i \rangle \simeq -g\gamma_i$, where $g = 2(1-\sigma_\epsilon^2)$ is the shear-ellipticity conversion factor, and $\sigma_\epsilon^2 = \langle \epsilon_1^2 \rangle = \langle \epsilon_2^2 \rangle$ is the variance of the intrinsic source ellipticities. Since we have $\gamma_1=\kappa \cos\alpha$ and $\gamma_2=\kappa \cos\beta$, $\langle \gamma^2 \rangle=\langle \gamma_1^2 \rangle+\langle \gamma_2^2 \rangle=\langle \kappa^2 \rangle$ and the above expression follows. 

Thus, 
\begin{equation} \label{ellipticity}
\epsilon ({\hat {\bf n}}) = 1.68 \times \frac{3}{2} \Omega_0 \left( \frac{H_0}{c}\right)^2 \int _0 ^{\tau _0} d \eta \frac{g(\eta)}{a(\tau_0 - \eta)} \int \frac{d ^3 {\bf k}}{(2 \pi)^{3/2}} D(\tau_0 - \eta) \delta ({\bf k}) e^{i {\bf k} \cdot {\hat {\bf n}} \eta}
\end{equation}

Now we perform the cross-correlation with the CMB maps for each of the three low-redshift effects described in \S~\ref{zcmb}.

\subsection{SZ Effect} \label{sz-cc}

The thermal SZ effect (\S~\ref{sz-desc}) in a given direction is computed as a line-integral. In the Rayleigh-Jeans portion of the spectrum (i.e. the long-wavelength limit) this is given by,
\begin{eqnarray} \label{sz-dt}
\lefteqn{\left[\frac{\Delta T}{T}({\hat {\bf m}})\right]_{SZ} \simeq \frac{-2 \sigma_T k_B b_{gas}}{m_e c^2}  \int_0^{\tau_0} d\eta' n_e(\eta') T_e(\eta') \delta ({\bf x'}) \nonumber} \\
&=& \frac{-2 \sigma_T k_B n_e^0 T_e^0 b_{gas}}{m_e c^2} \int_0^{\tau_0} d\eta' {a(\tau_0-\eta')}^{-\beta} \int \frac{d ^3 {\bf k'}}{(2 \pi)^{3/2}} D(\tau_0 - \eta') \delta ({\bf k'}) e^{i {\bf k'} \cdot {\hat {\bf m}} \eta'}
\end{eqnarray}
where $n_e^0$ and $T_e^0$ denote the electron number density and temperature at the present epoch, respectively, and $\beta$ is a parameter describing the redshift evolution of $n_e$ and $T_e$. For the calculations in this paper, we assume that $\beta=0$, i.e. that the pressure is independent of redshift. Note that $a(\tau_0-\eta') = 1/(1+z)$. $b_{gas}$ is the bias factor of the gas relative to the dark matter density perturbations $\delta ({\bf x'})$. Again, we have assumed a linear bias model with a constant bias factor for simplicity. In order to produce a conservative estimate of signal-to-noise for the SZ correlation, we have used $b_{gas}=4.0$, which is somewhat lower than the average bias predicted by numerical simulations for the low-$k$ regime (\cite{ref99}).

$n_e^0$ is calculated assuming $\Omega_b h^2 = 0.015$ where $h$ is the Hubble constant in units of 100 kms$^{-1}$ Mpc$^{-1}$. $T_e^0$  is taken to be 0.5 keV. $\sigma_T = 6.65 \times 10^{-25}$ cm$^2$ is the Thompson scattering cross-section, $c$ is the speed of light, and $m_e$ is the electron rest mass. 

Using Eqs.~\ref{counts-k} and~\ref{sz-dt}, the CMB-number count cross-correlation is,
\begin{eqnarray} \label{sz-cc1}
\left\langle \frac{\Delta T}{T}({\hat {\bf m}}) \Sigma _i ({\hat {\bf n}}) \right\rangle_{SZ} &=& \frac{2 \sigma_T k_B n_e^0 T_e^0 b_{gas}}{m_e c^2} \int _0 ^{\tau _0} d \eta \left[ b_i W(\eta) + 3 \Omega _0 \left( \frac{H_0}{c} \right)^2(2.5s_i -1) \frac{g(\eta)}{a(\tau_0 - \eta)} \right] \nonumber \\
&\times& D(\tau_0 - \eta) \int_0^{\tau_0} d\eta' \frac{D(\tau_0 - \eta')}{{a(\tau_0-\eta')}^{\beta}} \int \frac{dk k^2} {(2 \pi)^3} P(k) \int d\Omega_k e^{i {\bf k} \cdot {\hat {\bf n}} \eta} e^{-i {\bf k} \cdot {\hat {\bf m}} \eta'}
\end{eqnarray}
Here, we have used 
\begin{equation} \label{deldel}
\left \langle \delta ({\bf k}) \delta ({\bf k'}) \right\rangle = \delta_{Dirac}({\bf k} + {\bf k'}) P(k)
\end{equation}
where $P(k)$ is the mass power spectrum. Expanding out the exponentials yields,
\begin{equation} \label{expexp}
\int d\Omega_k e^{i {\bf k} \cdot {\hat {\bf n}} \eta} e^{-i {\bf k} \cdot {\hat {\bf m}} \eta'} = 4\pi \sum_{\ell} (2\ell+1) j_\ell(k\eta) j_\ell(k\eta') P_\ell({\hat {\bf m}} \cdot {\hat {\bf n}})
\end{equation}
Thus, we have
\begin{equation} \label{sz-cc-main}
\left\langle \frac{\Delta T}{T}({\hat {\bf m}}) \Sigma _i ({\hat {\bf n}}) \right\rangle_{SZ} = \sum_{\ell} \frac {(2\ell+1)}{4\pi} C_\ell^{\Sigma _i-SZ} P_\ell({\hat {\bf m}} \cdot {\hat {\bf n}}) 
\end{equation}
where
\begin{equation} \label{sz-counts-cl}
C_\ell^{\Sigma _i-SZ} = \frac{2}{\pi} \left( \frac{2 \sigma_T k_B n_e^0 T_e^0 b_{gas}}{m_e c^2} \right) \int k^2 dk P(k) w_\ell^{SZ}(k) w_\ell^{\Sigma _i}(k)
\end{equation}
with
\begin{equation} \label{wl-sigi}
w_\ell^{\Sigma _i}(k) = \int _0 ^{\tau _0} d \eta \left[ b_i W(\eta) + 3 \Omega _0 \left( \frac{H_0}{c} \right)^2(2.5s_i -1) \frac{g(\eta)}{a(\tau_0 - \eta)} \right] D(\tau_0 - \eta) j_\ell(k\eta)
\end{equation}
and
\begin{equation} \label{wl-sz}
w_\ell^{SZ}(k) = \int_0^{\tau_0} d\eta' \frac{D(\tau_0 - \eta')}{{a(\tau_0-\eta')}^{\beta}} j_\ell(k\eta')
\end{equation}
Note also that the number count autocorrelation is
\begin{equation} \label{counts-counts-cl}
C_\ell^{\Sigma _i-\Sigma _i} = \frac{2}{\pi} \int k^2 dk P(k) \left[ w_\ell^{\Sigma_i}(k) \right]^2
\end{equation}

The signal-to-noise per $\ell$-mode for a cross-correlation, assuming the correlation is weak, is $(S/N)_\ell^2 = \langle a b \rangle ^2 /(\langle a^2 \rangle \langle b^2 \rangle)$. Applying this formula to the SZ-number count cross-correlation yields,
\begin{equation} \label{sz-sn}
\left(\frac{S}{N}\right)_{SZ}^2 = \frac{\Delta\Omega}{4\pi} \sum_{\ell} (2\ell+1) \frac {(C_\ell^{\Sigma _i-SZ})^2}{(C_\ell^{\Sigma _i-\Sigma _i}+C_\ell^{POISSON})(C_\ell^{TT}+C_\ell^{DET})}
\end{equation}
where $C_\ell^{TT}$ is the total CMB signal (which is dominated by contributions from the surface of last scatter), $\Delta\Omega=\pi$ is the SDSS sky coverage, and the shot noise contribution per mode is,
\begin{equation} \label{poisson}
C_\ell^{POISSON} = \frac{1}{N_i}
\end{equation}
where $N_i$ is the number of tracer objects $i$ per steradian (\cite{pee80}). The detector contribution per $\ell$-mode is,
\begin{equation} \label{detnoise}
C_\ell^{DET} = n_0^2 ({w_\ell^{exp}})^{-2}
\end{equation}
where $w_\ell^{exp}$ is the experimental window function, and $n_0$ is a quantity denoting the noise level in $\delta T/T$ per pixel. We compute $C_\ell^{DET}$ by combining the 45, 60 and 90 GHz channels of the MAP experiment.

Next, we develop an optimal filter function which maximizes the signal-to-noise of the cross-correlation.

\subsubsection{Maximizing the signal} \label{max}

We want to find the appropriate weight function, $w_\ell(k,\eta)$, that maximizes the signal from $a_{lm}\tilde{a}_{lm}$ where,
\begin{equation}
a_{lm} = a_{lm}^{LS} + \int d\eta' \int k^2 dk A_{lm}(k) j_\ell(k\eta')f(\eta')
\end{equation}
with the first term in the sum being the contribution from the surface of last scatter, and
\begin{equation}
\tilde{a}_{lm} = \int d\eta \int k^2 dk A_{lm}(k) D(\tau_0-\eta) w_\ell(k,\eta)j_\ell(k\eta)
\end{equation}
with $A_{lm}(k)$ as the spherical harmonic expansion of the comoving number density,
\begin{equation}
n_i(x) = \int k^2 dk A_{lm}(k) D(\tau_0-\eta) j_\ell(k\eta) Y_{lm}(\Omega)
\end{equation}
and
\begin{equation}
f(\eta') = \frac{2 \sigma_T k_B n_e^0 T_e^0 b_{gas}}{m_e c^2} \frac{D(\tau_0 - \eta')}{{a(\tau_0-\eta')}^{\beta}}
\end{equation}

The noise term is
\begin{equation}
\left\langle [A_{lm}(k) D(\tau_0-\eta)][A_{lm}^{\ast}(k) D(\tau_0-\eta')]\right\rangle = b_i^2 P(k) D(\tau_0-\eta) D(\tau_0-\eta') + n_i^{-1/2}(\eta) n_i^{-1/2}(\eta')
\end{equation}
where $n_i(\eta)$ depends on the galaxy selection function. In evaluating this function, space is considered split into a number of discrete time regions. We want to maximize
\begin{eqnarray}
\chi^2_{lm} &=& \frac{{\left\langle a_{lm}\tilde{a}_{lm} \right\rangle}^2}{{\left\langle a^2_{lm} \right\rangle}{\left\langle \tilde{a}^2_{lm} \right\rangle}} \nonumber \\
&=& \frac{\left[\sum_{{\bf k}} k^2 \int \int d\eta d\eta' w_\ell(k,\eta) j_\ell(k\eta) j_\ell(k\eta') f(\eta') b_i^2 P(k) D(\eta) \right]^2}{{\left\langle a^2_{lm} \right\rangle}{\sum_{{\bf k}}\int \int d\eta d\eta' w_\ell(k,\eta) w_\ell(k,\eta') j_\ell(k\eta) j_\ell(k\eta') [b_i^2 P(k) D(\eta) D(\eta') + n_i^{-1/2}(\eta) n_i^{-1/2}(\eta')]}} \nonumber \\
 & & 
\end{eqnarray}
Differentiating $\ln \chi^2_{lm}$ with respect to $w_\ell(k,\eta)$ and setting to zero yields:
\begin{eqnarray}
\lefteqn{\frac{k^2 D(\eta) j_\ell(k\eta) b_i^2 P(k) \int d\eta'  j_\ell(k\eta') f(\eta')}{\sum_{{\bf k}} k^2 \int \int d\eta d\eta' w_\ell(k,\eta) j_\ell(k\eta) j_\ell(k\eta') f(\eta') b_i^2 P(k) D(\eta)} \nonumber} \\
&=& \frac{k^2 j_\ell(k\eta) [b_i^2 P(k) D(\eta) \int d\eta' j_\ell(k\eta') D(\eta') w_\ell(k,\eta') + n_i^{-1}(\eta) \int d\eta' j_\ell(k\eta') w_\ell(k,\eta')]}{\sum_{{\bf k}}\int \int d\eta d\eta' w_\ell(k,\eta) w_\ell(k,\eta') j_\ell(k\eta) j_\ell(k\eta') [b_i^2 P(k) D(\eta) D(\eta') + n_i^{-1/2}(\eta) n_i^{-1/2}(\eta')]} \nonumber \\
& &
\end{eqnarray}
If we ignore the slow variations in the counts and growth factors, then the solution to this equation is
\begin{equation} \label{opt-filter}
w_\ell(k,\eta') = \frac{b_i^2 P(k)f(\eta')}{b_i^2 P(k)D(\tau_0-\eta') + n_i^{-1}(\eta')/D(\tau_0-\eta')}
\end{equation}
This weighting is just the projected signal over the projected noise. By weighting the integrand of Eq.~\ref{sz-counts-cl} with this optimal filter, one of the $C_\ell^{\Sigma _i-SZ}$ factors in the numerator of Eq.~\ref{sz-sn} cancels with the $(C_\ell^{\Sigma _i-\Sigma _i}+C_\ell^{POISSON})$ factor in the denominator, and we are left with:
\begin{equation} \label{sz-sn-opt}
\left(\frac{S}{N}\right)_{SZ}^2 = \frac{\Delta\Omega}{4\pi} \sum_{\ell} (2\ell+1) \frac {\left|C_\ell^{\tilde{\Sigma}_i-SZ}\right|}{C_\ell^{TT}+C_\ell^{DET}}.
\end{equation}

Prior to this point, the linear power spectrum $P(k)$ was used in order to make the derivations of the cross-correlation and the optimal filter clearer. However, in the actual calculations we will use the fully-evolved non-linear power spectrum. In the non-linear case, the growth factor $D$ gets absorbed into the power spectrum, which becomes a function of both wavenumber $k$ and scale factor $a$: $P(k,a)$. 

Let us define the quantity
\begin{equation} \label{sqrtP}
\wp(k,a) = \sqrt{P(k,a)}.
\end{equation}

Using this non-linear power spectrum, the quantities in Eq.~\ref{sz-sn-opt} are given by,
\begin{equation} \label{sz-opt-cl}
C_\ell^{\tilde{\Sigma}_i-SZ} = \frac{2}{\pi} \left( \frac{2 \sigma_T k_B n_e^0 T_e^0 b_{gas}}{m_e c^2} \right)^2 \int k^2 dk w_\ell^{SZ}(k) w_\ell^{\tilde{\Sigma}_i-SZ}(k),
\end{equation}
with 
\begin{equation} \label{wl-opt-sz}
w_\ell^{\tilde{\Sigma}_i-SZ}(k) = \int_0^{\tau_0} d\eta \frac{\wp(k,a)}{{a(\tau_0-\eta)}^{\beta}} j_\ell(k\eta) \left[ 1-\frac{1}{1+b_i^2 P(k,a) n_i(\eta)} \right]
\end{equation}
and Eq.~\ref{wl-sz} becomes
\begin{equation} \label{wl-sz-nl}
w_\ell^{SZ}(k) = \int_0^{\tau_0} d\eta' \frac{\wp(k,a')}{{a(\tau_0-\eta')}^{\beta}} j_\ell(k\eta').
\end{equation}

Note that in the case of the SZ-ellipticity correlation for the FIRST objects, the signal-to-noise is defined analogous to Eq.~\ref{sz-sn} with
\begin{equation} \label{sz-e-cl}
C_\ell^{\epsilon-SZ} = \frac{2}{\pi} \left( \frac{2 \sigma_T k_B n_e^0 T_e^0 b_{gas}}{m_e c^2} \right) \int k^2 dk w_\ell^{SZ}(k) w_\ell^{\epsilon}(k)
\end{equation}
where
\begin{equation} \label{wl-e}
w_\ell^{\epsilon}(k) = \int _0 ^{\tau _0} d \eta \left[ 1.68 \times \frac{3}{2} \Omega_0 \left( \frac{H_0}{c}\right)^2  \frac{g(\eta)}{a(\tau_0 - \eta)} \right] \wp(k,a) j_\ell(k\eta)
\end{equation}
Also,
\begin{equation} \label{e-e-cl}
C_\ell^{\epsilon-\epsilon} = \frac{2}{\pi} \int k^2 dk \left[ w_\ell^\epsilon (k) \right]^2
\end{equation}
and
\begin{equation} \label{poisson-e}
C_\ell^{POISSON} = \frac{\sigma_\epsilon^2}{N_{\mathit{FIRST}}}
\end{equation}
where $\sigma_\epsilon = 0.4$ as before, and $N_{\mathit{FIRST}}$ is the number of FIRST objects per steradian. We estimate $\sim$10 FIRST sources per square degree will be optically matched with the SDSS. 

\subsection{ISW Effect} \label{isw-cc}

The temperature fluctuations arising from the ISW effect can be expressed as,
\begin{equation} \label{isw-basic}
\left[\frac{\Delta T}{T}\right]_{ISW} =\frac{2}{c^2} \int d\eta \frac{\partial \Phi}{\partial \eta}
\end{equation}
We can relate the gravitational potential $\Phi$ to the comoving density field $\delta$:
\begin{equation} \label{del-potential}
\nabla^2\Phi({\bf x}) = \frac{4 \pi G \rho_0}{a} \delta({\bf x})
\end{equation}
where the Laplacian is evaluated in comoving coordinates. Expanding $\delta({\bf x})$ using Eq.~\ref{delta-expand}, we obtain 
\begin{equation} \label{potential}
\Phi = 4 \pi G \rho_0 \frac{D(\tau_0-\eta')}{a(\tau_0-\eta')} \int \frac{d ^3 {\bf k}}{(2 \pi)^{3/2}} \frac{\delta ({\bf k})} {k^2} e^{i {\bf k} \cdot {\hat {\bf m}} \eta'}
\end{equation}
Combining Eqs.~\ref{isw-basic} and ~\ref{potential}, the ISW term becomes
\begin{equation}
\left[ \frac{\Delta T}{T}({\hat {\bf m}}) \right]_{ISW} = \frac{3 H_0^2 \Omega_0}{c^2} \int_0^{\tau_0} d\eta' \frac{\partial}{\partial \eta'} \left[\frac{D(\tau_0-\eta')}{a(\tau_0-\eta')}\right] \int \frac{d ^3 {\bf k}}{(2 \pi)^{3/2}} \frac{\delta ({\bf k})} {k^2} e^{i {\bf k} \cdot {\hat {\bf m}} \eta'}
\end{equation}
where we have used $8 \pi G \rho_0 = 3 H_0^2 \Omega_0$. Following the same formalism as in \S~\ref{max} to derive an optimal filter for the ISW-number-count cross-correlation, we obtain
\begin{equation} \label{isw-opt-cl}
C_\ell^{\tilde{\Sigma}_i-ISW} = \frac{2}{\pi} \left( \frac{3 H_0^2 \Omega_0}{c^2} \right)^2 \int \frac{dk}{k^2} w_\ell^{ISW}(k) w_\ell^{\tilde{\Sigma}_i-ISW}(k)
\end{equation}
where 
\begin{equation} \label{wl-isw}
w_\ell^{ISW} = \int_0^{\tau_0} d\eta' \frac{\partial}{\partial \eta'} \left[\frac{\wp(k,a')}{a(\tau_0-\eta')}\right] j_\ell(k\eta')
\end{equation}
and
\begin{equation} \label{wl-opt-isw}
w_\ell^{\tilde{\Sigma}_i-ISW}(k) = \int_0^{\tau_0} d\eta \frac{\partial}{\partial \eta} \left[\frac{\wp(k,a)}{a(\tau_0-\eta)}\right] j_\ell(k\eta) \left[ 1-\frac{1}{1+b_i^2 P(k,a) n_i(\eta)} \right]
\end{equation}
Hence, following the notation of Eq.~\ref{sz-sn}, the signal-to-noise is:
\begin{equation} \label{isw-sn-opt}
\left(\frac{S}{N}\right)_{ISW}^2 = \frac{\Delta\Omega}{4\pi} \sum_{\ell} (2\ell+1) \frac {\left| C_\ell^{\tilde{\Sigma}_i-ISW}\right|}{C_\ell^{TT}}
\end{equation}
We have neglected $C_\ell^{DET}$ in the denominator since the ISW effect is confined to low $\ell$ where MAP's detector noise is very low.

For FIRST objects, the signal-to-noise is
\begin{equation} \label{isw-e-sn}
\left(\frac{S}{N}\right)_{ISW}^2 = \frac{\Delta\Omega}{4\pi} \sum_{\ell} (2\ell+1) \frac {(C_\ell^{\Sigma_i-ISW})^2}{C_\ell^{TT}(C_\ell^{\epsilon-\epsilon} + C_\ell^{POISSON})}
\end{equation}
where
\begin{equation} \label{isw-e-cl}
C_\ell^{\epsilon-ISW} = \frac{2}{\pi} \left( \frac{3 H_0^2 \Omega_0}{c^2} \right) \int dk w_\ell^{ISW}(k) w_\ell^{\epsilon}(k)
\end{equation}
and the other $C_\ell$ and $w_\ell$ terms are as defined previously in Eqs.~\ref{wl-e},~\ref{e-e-cl},~\ref{poisson-e} and ~\ref{wl-isw}.

\subsection{Weak Lensing} \label{lensing-cc}

We follow the methodology and notation of \cite{zal99}. Most tracers of the underlying density field will not be perfect since they correlate only partially with the convergence $\kappa$. Assuming we have a map $Y$ that correlates with $\kappa$ and has a cross-correlation $C_\ell^{\kappa Y}$, the cross-correlation between an $\varepsilon$ map and the $Y$ map gives signal-to-noise
\begin{equation} \label{lensing-sn}
\left(\frac{S}{N}\right)_{\mathit{lensing}}^2 = \frac{\Delta\Omega}{4\pi} \sum_{\ell} (2\ell+1) \frac {4 {W_{res}(\ell)}^2 (C_\ell^{\kappa Y})^2}{(4 {W_{res}(\ell)}^2 C_\ell^{\kappa\kappa}+N_\ell^{\varepsilon\varepsilon})(C_\ell^{YY}+C_\ell^{POISSON})}
\end{equation}
where $\varepsilon$ is a spin-zero quantity charactering the CMB polarization field. In this expression, the window function describing the effect of finite angular resolution is,
\begin{equation}
W_{res}(q) = \frac{\int_0^{2\pi} d\theta \int_0^{\ell_{cut}} \ell d\ell F(\ell) F(\sqrt{\ell^2 + q^2 + 2 \ell q \cos \theta}) \ell^2 C_\ell^{TT}}{2 \pi \int_0^{\ell_{cut}} \ell d\ell {F(\ell)}^2 C_\ell^{TT}}
\end{equation}
where $C_\ell^{TT}$ is the CMB power spectrum, and
\[ \mbox{$F(\ell)$} = \left\{ \begin{array}{ll}
	1 & \mbox{$\ell < \ell_{cut}$} \\
	0 & \mbox{$\ell > \ell_{cut}$}
	\end{array}
	      \right. \]
For the MAP experiment, $\ell_{cut} = 600$. Also, the large-scale amplitude of the noise spectrum is, 
\begin{equation}
N_\ell^{\varepsilon\varepsilon} = \frac{ 2 \pi \int_0^{\ell_{cut}} \ell d\ell \ell^4 {F(\ell)}^4 (C_\ell^{TT} + C_\ell^{DET})}{(\int \ell d\ell \ell^2 {F(\ell)}^2 C_\ell^{TT})^2}
\end{equation}
The convergence autocorrelation is given by
\begin{equation} \label{kk-cl}
C_\ell^{\kappa\kappa} = \frac{2}{\pi} \left( \frac{3 H_0^2 \Omega_0}{2 c^2} \right)^2 \int k^2 dk \left[ w_\ell^{\kappa}(k)\right]^2
\end{equation}
where 
\begin{equation} \label{wl-kap}
w_\ell^{\kappa} = \int_0^{\tau_0} d\eta' \frac{\wp(k,a')}{a(\tau_0-\eta')} g(\eta') j_\ell(k\eta')
\end{equation}
If the $Y$ map consists of ellipticities of FIRST objects, then $C_\ell^{YY} = C_\ell^{\epsilon\epsilon}$, and $C_\ell^{\kappa Y}$ becomes
\begin{equation} \label{kY-e-cl}
C_\ell^{\kappa\epsilon} = \frac{2}{\pi} \left( \frac{3 H_0^2 \Omega_0}{2 c^2} \right) \int k^2 dk w_\ell^{\kappa}(k) w_\ell^{\epsilon}(k)
\end{equation}
where $w_\ell^{\epsilon}$ is from Eq.~\ref{wl-e}.

If the $Y$ map consists of a number counts of tracer objects $i$, then we can use the optimal filter technique explained previously. Thus, the signal-to-noise becomes,
\begin{equation} \label{lensing-opt-sn}
\left(\frac{S}{N}\right)_{\mathit{lensing}}^2 = \frac{\Delta\Omega}{4\pi} \sum_{\ell} (2\ell+1) \frac {4 {W_{res}(\ell)}^2 \left| C_\ell^{\tilde{\Sigma}_i-\kappa}\right|}{4 {W_{res}(\ell)}^2 C_\ell^{\kappa\kappa}+N_\ell^{\varepsilon\varepsilon}}
\end{equation}
where
\begin{equation} \label{kappa-opt-cl}
C_\ell^{\tilde{\Sigma}_i-\kappa} = \frac{2}{\pi} \left( \frac{3 H_0^2 \Omega_0}{c^2} \right)^2 \int k^2 dk w_\ell^{\kappa}(k) w_\ell^{\tilde{\Sigma}_i-\kappa}(k)
\end{equation}
and
\begin{equation} \label{wl-kappa-opt}
w_\ell^{\tilde{\Sigma}_i-\kappa} = \int_0^{\tau_0} d\eta' \frac{\wp(k,a')}{a(\tau_0-\eta')} g(\eta') j_\ell(k\eta') \left[ 1-\frac{1}{1+b_i^2 P(k,a') n_i(\eta')} \right]
\end{equation}

\subsection{Cross-correlation Computation} \label{cc-comp}

In evaluating the above integrals, the non-linear power spectrum $P(k,a)$ was computed using the analytic approximation for a fully-evolved mass power spectrum for spatially flat cold dark matter cosmological models with quintessence given by \cite{cpm99}. The linear power spectrum used in this approximation was computed using CMBFAST (\cite{slz96}), using adiabatic initial conditions and a scale-invariant primordial spectrum. Going to a non-linear power spectrum makes very little difference to the cross-correlations, since most power arises from the linear regime, $k\ll 1$ ($h$ Mpc$^{-1}$), at which the non-linear contribution to the spectrum starts to dominate.

To increase the speed of the numerical integrations, the $k$-integrals were done in logarithmic space and the spherical Bessel functions $j_l(k\eta)$ in the $\eta$-integrals were approximated by: 
\begin{equation} \label{limber}
j_{\ell}(k\eta) = \left\{ \begin{array}{l@{\quad:\quad}l}
	j_{\ell}(k\eta) & k\tau_0<\ell \\ 
	\sqrt{\frac{\pi}{2\ell}} \delta(k\eta-\ell) & k\tau_0>\ell
	\end{array} \right. 
\end{equation}

The optimal filter technique was applied to galaxies (both the photo-z and complete samples) and to quasars with photometric redshifts. The number density functions $n_i(\eta)$ used in the number count optimal filters were computed as follows: 

\begin{enumerate}

\item{The galaxy number density was calculated using the $r$-band Schecter-form luminosity function (LF) of galaxies in the Las Campanas Redshift Survey (LCRS) (\cite{lin96}).}

\item{For photometric ($i'<21$) quasars, the quasar luminosity function described in \cite{fan99} was used. This LF is defined for $h=0.5$, where $h \equiv H_0 /100$ km s$^{-1}$ Mpc$^{-1}$. Dependence on $h$ was added to the LF using the fact that $h^{-3} \Phi(M,z)$ and $M - 5 \log h$ are invariant under transformation of $h$. The LF was taken to be valid for absolute magnitudes $-30.0 \leq M-5 \log (h/0.5) \leq -23.0$. Finally, this LF is defined in the B band. To convert apparent magnitudes in the SDSS $i'$ band to $M_B$, we have followed \cite{sch91}:
\begin{equation} \label{i-to-B}
M_B = i' - 5 \log \left[\eta(1+z)\right] - 25.0 + 2.5 \log (1+z) - 2.5 \alpha \log \left[ \frac{7628}{(1+z)4400}\right] + 0.15
\end{equation}
}
Here, $\eta(1+z)$ is the luminosity distance (in Mpc); $\alpha=-0.5$ is the power law index of the quasar spectrum; the $i'$ and $B$ bands have effective wavelengths of 7628 and 4400 \AA~respectively, and the offset between the $B$ and the $AB$ magnitude systems is 0.15 (\cite{fuk96}).
\end{enumerate}

The comoving number density is given by,
\begin{equation} \label{num-den}
n(z) = (1+z)^3 \int_{M_{min}(z)}^{M_{max}(z)} \Phi(M, z) dM
\end{equation}
where $\Phi(M, z)$ is the LF.

The cross-correlations of spectrometric quasars and FIRST objects were calculated using the non-filtered formalism (Eqs.~\ref{sz-sn},~\ref{isw-e-sn} and~\ref{lensing-sn}). The radial distribution $W(\eta)$ of spectrometric ($i'<19$) quasars was calculated using the redshifts of a set of quasars simulated in SDSS colors supplied by Donald Schneider. The radial distribution of FIRST objects was approximated by a Gaussian with mean $z=1.2$ and width $\Delta z = 0.4$.


\section{CROSS-CORRELATION RESULTS} \label{cc-res}

Table~\ref{tab1} shows the estimated $(S/N)^2$ that would be obtained by cross-correlating MAP with the SDSS survey for a pair of models with constant $\Omega_0 h^2$. 
\placetable{tab1}

The SZ and weak lensing cross-correlations give interesting values of signal-to-noise and can be used in various ways, described below.

\subsection{Tracing Gas Pressure Fluctuations} \label{sz-res}

While clusters produce the highest amplitude Sunyaev-Zeldovich fluctuations, large scale variations in density associated with the superclusters are likely the dominant source of large angle SZ fluctuations (\cite{per95}; \cite{ref99}). We can use the SDSS photo-z galaxy sample ($r'<21$) to trace the large-scale structure out to $z \sim 0.5$. By measuring the cross-correlation between the MAP CMB measurements and the galaxies in each photo-z slice, we can measure the amplitude of the gas pressure/galaxy cross-correlation in each photo-z redshift bin. 

For the SZ cross-correlation, we have from Eq.~\ref{sz-sn-opt} that $(S/N)^2 \propto b_{gas} T_e^0 $. The gas is highly biased ($b_{gas} \sim 4$, \cite{ref99}), and is at high temperature ($T_e \sim 0.5$ keV). Since the surface number density of the photo-z galaxies is very high, the Poisson noise in the cross-correlation is minimized. Furthermore, the SZ signal is weighted by the growth factor but not by the scale factor for $\beta=0$ (Eq.~\ref{wl-opt-sz}). Therefore most of the signal arises from low-redshift space where the galaxy density is highest. These factors contribute to a high signal-to-noise from this cross-correlation. 

\begin{figure} \plotone{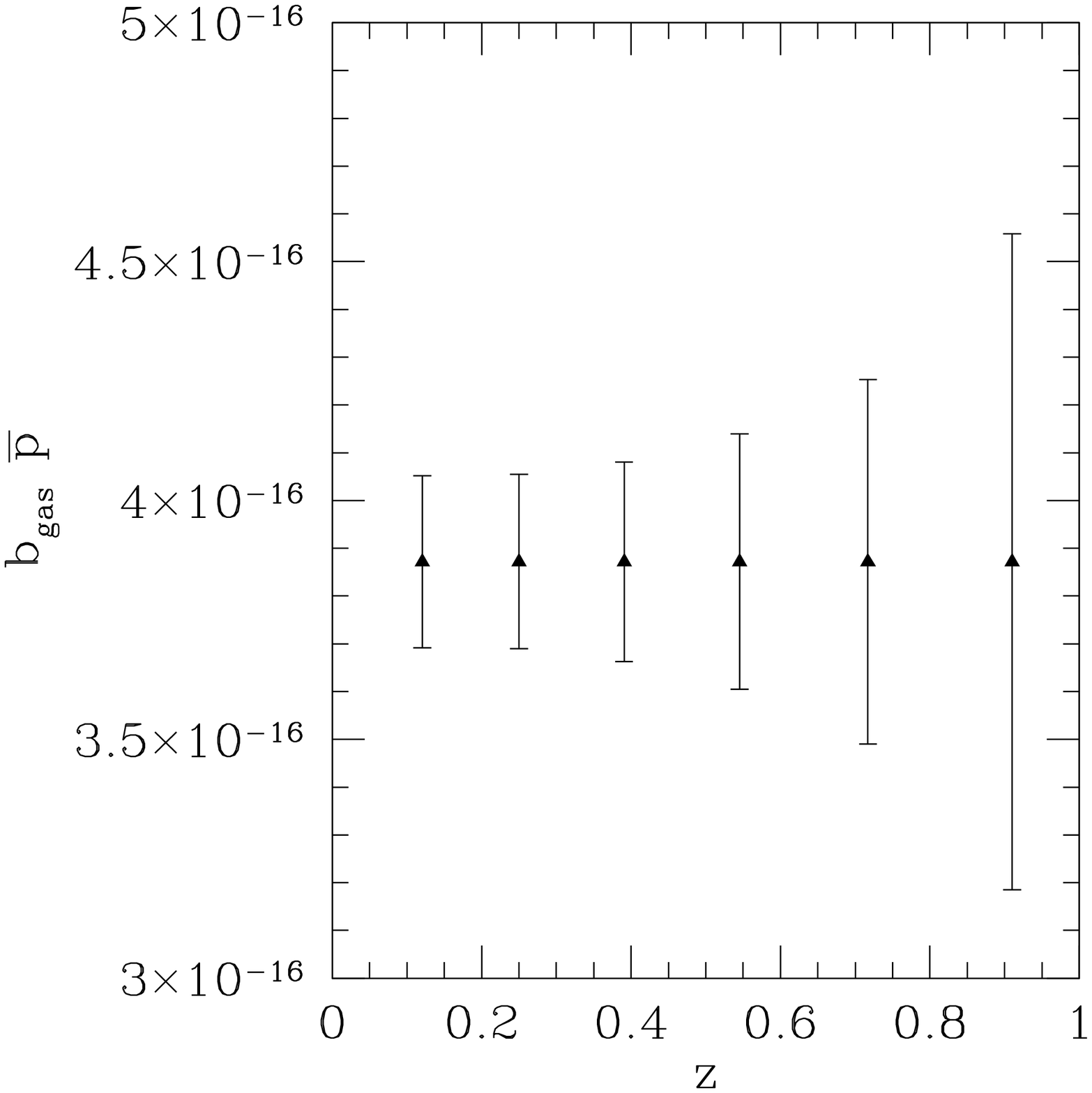} \caption{{\bf Pressure Fluctuations} This figure shows the predicted measurement of the comoving pressure fluctuations based on measurements of the correlation between photometrically-selected galaxies ($r'<21$), separated into redshift ``slices'', and CMB temperature fluctuations (see \S~\ref{sz-res}). Pressure is in cgs units. The calculation assumes that $T_e = 0.5$ keV today and $b_{gas} = 4$. 
\label{fig1}} \end{figure}

Fig.~\ref{fig1} shows our predicted signal and error in our measurements of $b_{gas} \overline{p}_{gas}$, where $b_{gas}$ is the ratio of the pressure fluctuations to the density fluctuations and $\overline{p}_{gas}$ is the mean gas pressure. The points for $b_{gas} \overline{p}$ are simply $1.936 \times 10^{-16}~a^{-\beta}$ ergs cm$^{-3}$, which is just a constant when $\beta=0$ as we have here. This calculation illustrates how well the deviations from the average pressure can be constrained. The error bars are calculated from the inverse of the SNR, which is obtained from Eq.~\ref{sz-sn}, but with the tracer galaxies selected from a redshift "slice". This is done by multiplying the radial distribution function by a gaussian of the required slice width, centered on the average redshift of that slice. Since most of the power in the cross-correlation arises in the linear
regime, the covariance between redshift bins should not have a significant effect.

These observations will yield the thermal history of the universe and a detection of the ``missing baryons" (\cite{cen99}, L. A. Phillips et al. 2000, in preparation).

\placefigure{fig1}

\subsection{Lensing and the CMB} \label{lensing-res}

The weak-lensing signal is inherently small, so it is important to choose the low-z tracer carefully in order to maximize the SNR. Many different effects are important in determining how well a given class of low-z tracers will perform when cross-correlated with weak lensing of the CMB.

Quasars are useful tracers of large scale density field. They are thought to be biased tracers of the underlying mass distribution.  Since the quasar number counts are steep ($d\ln N/d m \sim 1$) for $i' < 19$, gravitational lensing by the foreground matter should produce significant fluctuations in the observed quasar angular correlations. These two different effects (bias and steep number-count slope) are important on different angular scales. Quasars can also probe the density field out to high redshifts ($z \sim 4$), thus sampling a significant fraction of the volume of redshift space. However, their surface number density is low, so depending on the cut used to select the quasars, the Poisson error can be large.

The surface number density of galaxies is very high, so the Poisson noise of the sample is small. However, the slope of the galaxy number counts is shallow ($d\ln N/d m \sim 0.5$ in $r'$ band) so the intrinsic fluctuation term of Eq.~\ref{counts-x} dominates over the lensing term and the cross-correlation with weak lensing does not trace the foreground matter fluctuations very well. The galaxy sample only extends out to $z \sim 0.5$, and thus probes only a small fraction of the volume probed by quasars.

We find that cross-correlation of weak lensing with $i'<19$ quasars in the SDSS survey is Poisson-noise-limited and does not provide very good signal-to-noise. The $r'<21$ galaxy sample performs better, but is hampered by the fact that it does not extend out to large redshifts. Filtering $i'<21$ photometric quasars, which have surface number density more than six times that of $i'<19$ quasars (Fan 1999), with the optimal filter developed in \S~\ref{max} gives reasonably high SNR. We will discuss how to use this cross-correlation to constrain $\Omega_0$ and the equation of state in \S~\ref{quin}.

\subsection{Cross-correlations with Poor SNR} \label{isw-res}

The ISW effect does not produce a good SNR, because it dominates at low $\ell$ where the matter power spectrum is falling and also because the SDSS redshift survey does not probe to sufficient depth. Optically matched FIRST objects in the SDSS survey are Poisson-noise limited and are found to be useless for the purpose of cross-correlating with the CMB. 

\section{EXPLORING THE EQUATION OF STATE} \label{quin}

The general equation of state of a density field $\rho$ is
\begin{equation} \label{eqstate}
\rho = \rho_0 a^{-3(1+w)} 
\end{equation}
where $w$ is the ratio of pressure to density. For example, $w=0$ for dark matter, $w=+1/3$ for radiation, and $w=-1$ for $\Lambda$. An alternative to the cosmological constant $\Lambda$ as the missing energy in a flat universe with $\Omega_0<1$ is given by quintessence $Q$ (\cite{cal98}). It is a time-evolving, spatially inhomogeneous energy component with negative pressure and an equation of state $w<0$. It drives cosmological expansion at late times, influencing the growth of structure. On very large scales, the quintessence clusters gravitationally, thereby modifying the level of CMB temperature anisotropy relative to the matter power spectrum amplitude and also inducing a late-time ISW effect. On small length scales, fluctuations in $Q$ disperse relativistically and the $Q$-field behaves as a smooth component. Quintessence modifies the matter power spectrum, the time evolution of the scale factor $a$ and the linear growth factor $D$. In this section we investigate the possibility of detecting observational imprints of $Q$ by cross-correlating MAP and SDSS.

Following the notation of \S~\ref{lensing-cc}, the $\chi^2$ for cross-correlating weakly-lensed $i'<21$ quasars with the CMB convergence is given by, 

\begin{eqnarray} \label{chi-sq-3d}
\chi^2 (\Omega_0, w, b_q) &=& \frac{\Delta\Omega}{4\pi} \sum_{\ell} (2\ell+1) \frac { \left[C_\ell^{\tilde{\Sigma}_q-\kappa}(\Omega_0, w, b_q) - C_\ell^{\tilde{\Sigma}_q-\kappa}(\Omega_0', w', b_q')\right]^2} {\left( C_\ell^{\kappa\kappa}+N_\ell^{\varepsilon\varepsilon}/{4 {W(\ell)}^2}\right)C_\ell^{\tilde{\Sigma}_q-\kappa}(\Omega_0', w', b_q')} \nonumber \\ 
&+& \frac{\Delta\Omega}{4\pi} \sum_{\ell} (2\ell+1) \frac { \left[C_\ell^{{\Sigma}_q-{\Sigma}_q}(\Omega_0, w, b_q) - C_\ell^{{\Sigma}_q-{\Sigma}_q}(\Omega_0', w', b_q')\right]^2}{\left({C_\ell^{POISSON}}\right)^2}
\end{eqnarray}

\begin{figure} \plotone{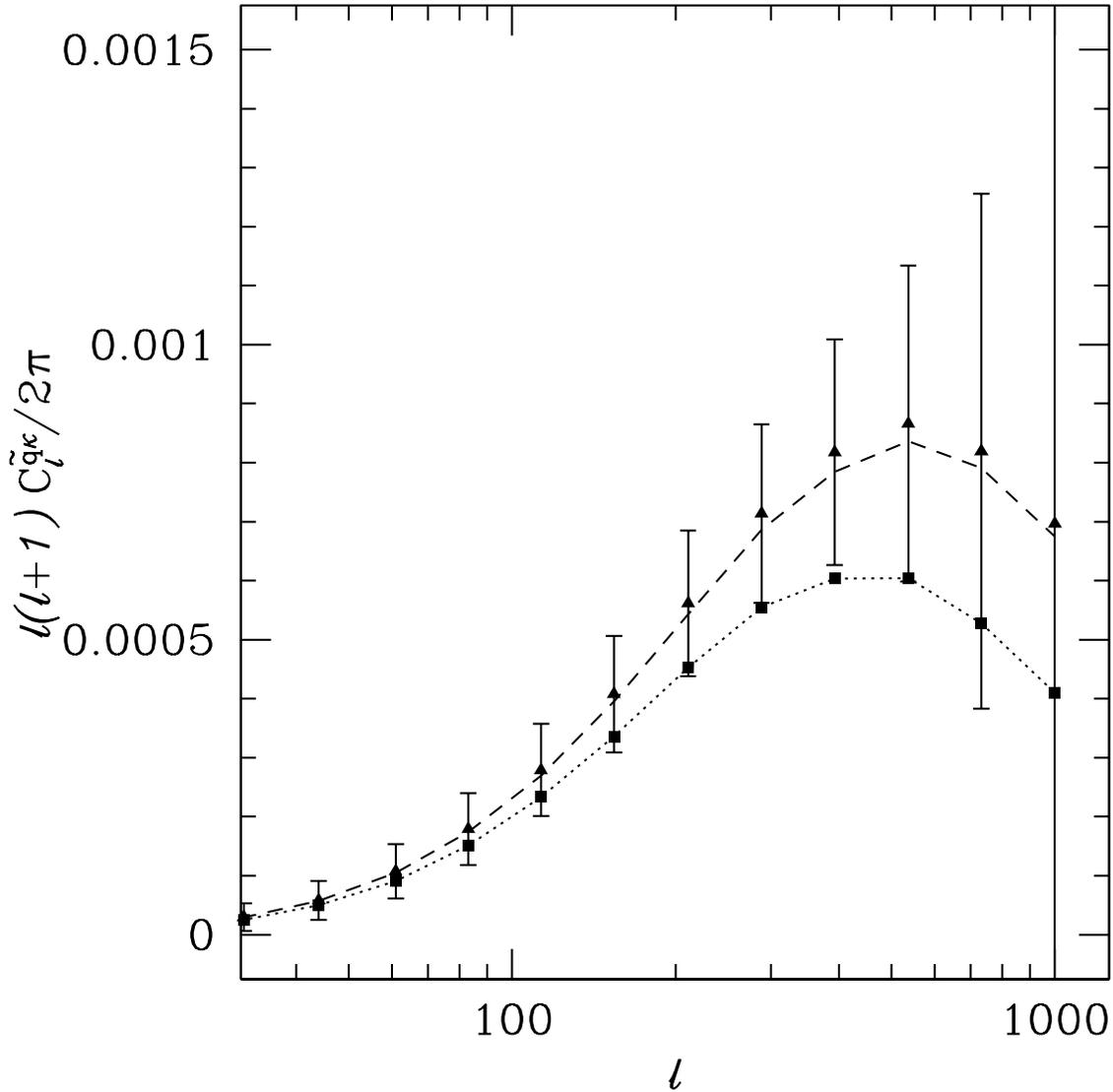} \caption{{\bf Quasar-Lensing Cross-Correlation} This figure shows the predicted correlation between $i'<21$ quasars in the SDSS survey and the convergence computed from the CMB sky (see \S~\ref{lensing-res}).  The solid lines are for a $\Lambda$CDM model with $\Omega_0 = 0.35$, $\Lambda = 0.65$, $H_0 = 70$, and $b_q = 2$.  The error bars are based on two year MAP data and the full SDSS survey. The dotted lines are a model with $\Omega_0 = 0.3$, $\Lambda = 0.7$, $H_0 = 76$ and $b_q = 2$. The dashed lines are for the same model with $b_q = 3$. The degeneracy of the parameters $\Omega_0$ and $b_q$ is apparent.
\label{fig2}} \end{figure}

We take the fiducial models $C_\ell^{\tilde{\Sigma}_q-\kappa}(\Omega_0', w', b_q')$ and $C_\ell^{{\Sigma}_q-{\Sigma}_q}(\Omega_0', w', b_q')$ to have $[\Omega_0=0.35, h^2=0.7, w=-1.0, b_q=3.0]$. It is seen from Figure~\ref{fig2} that there is a degeneracy between the quasar bias $b_q$ and $\Omega_0$. For each model, we minimize $\chi^2$ with respect to $b_q$.

\placefigure{fig2}

\begin{figure} \plotone{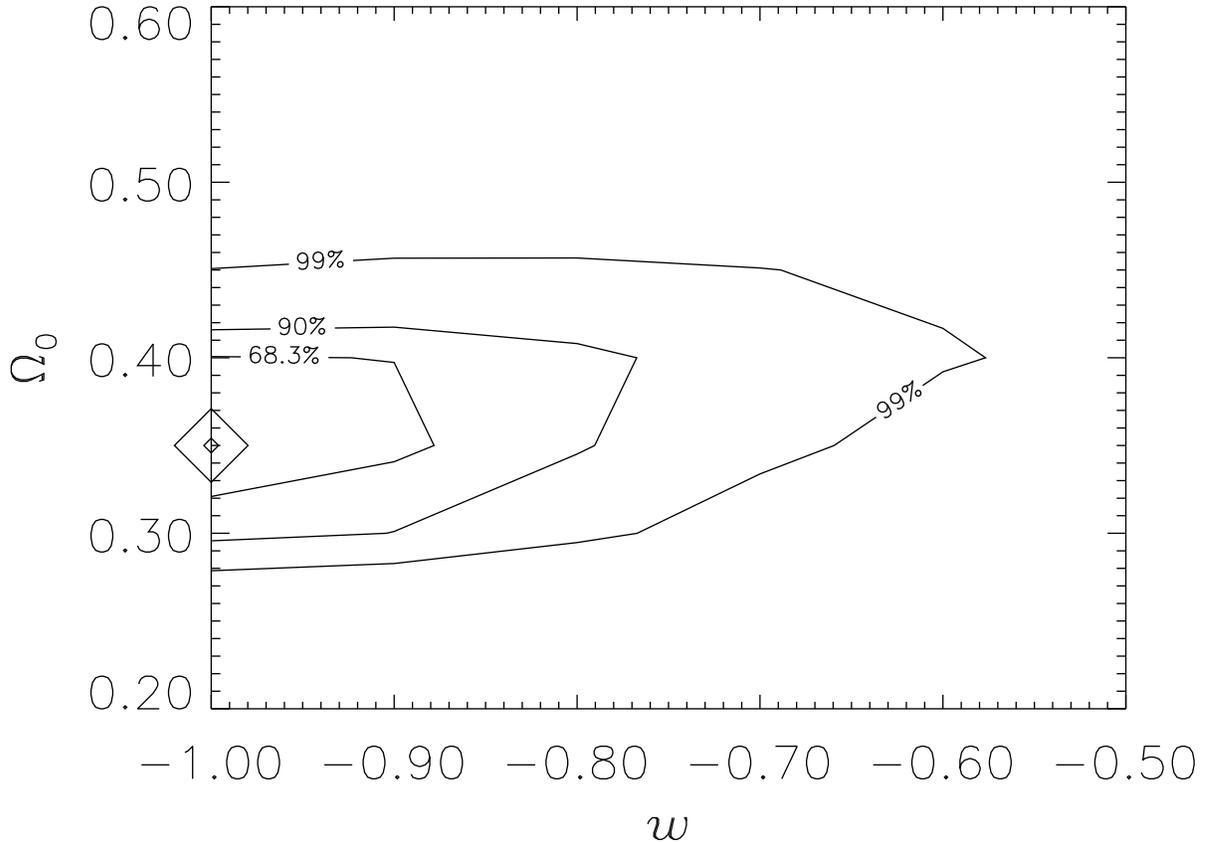} \caption{ {\bf Parameter Estimation} This figure shows the $\chi ^2$ surface for the $\Omega_0 - w$ parameter space of a family of spatially flat models with constant $\Omega_0 h^2$. These constraints are due to cross-correlation of weak lensing of quasar number counts for $i'<21$ with MAP CMB data. For each model, $\chi ^2$ has been minimized with respect to the quasar bias $b_q$. The fiducial model $[\Omega_0=0.35, h^2=0.70, w=-1.0, b_q=3.0]$ is indicated by the diamond. The curves project to one-dimensional intervals containing the probabilities with which they are labeled. \label{fig3}} \end{figure}

Exploring the parameter space for a family of spatially flat models with constant $\Omega_0 h^2$ for weak lensing of $i'<21$ quasar number counts, we obtain the likelihood surface shown in Figure~\ref{fig3}. Here, we have assumed that there are 60 $i'<21$ quasars/sq. deg., in accordance with~\cite{fan99}. The labeled curves project to one-dimensional intervals containing $68.3\%$, $90\%$ and $99\%$ of normally-distributed data.  

At $90\%$ probability, we obtain the following bounds: $0.30 \leq \Omega_0 \leq 0.42$ and $-1.0\leq w \leq -0.77$.

\placefigure{fig3}


\section{SUMMARY} \label{finish}

We apply a formalism for computing the cross-correlation between data-sets at high- and low-redshift to MAP data and various classes of tracer objects from the SDSS, for a range of spatially-flat low-$\Omega_0$ models with a general equation of state. Three low-z effects that imprint secondary anisotropies on the CMB are considered: the thermal Sunyaev-Zel'dovich effect, the Integrated Sachs-Wolfe effect and weak lensing. 

Using photometric redshifts, one can select a tracer sample to match the redshift windows where a particular second-order CMB anisotropy is dominantly formed, using the optimal filtering technique detailed here. As a practical algorithm, one would derive the luminosity function of the tracer sample, and hence the comoving number density of that sample for a particular cosmology. Then one would guess a power spectrum and tracer bias, making use of the CMB and tracer power spectra; construct the optimal weighting for each 3-D Legendre mode as a function of redshift and multipole; and finally project to form a two-dimensional map which is then cross-correlated with the CMB. One then estimates the power spectrum and bias from this map, and repeats the procedure until the process converges to the real optimal filter.

We find that these techniques would give good constraints on pressure fluctuations of supercluster-scale gas and provide constraints on $\Omega_0$ and the equation of state $w$ in quintessence models. It is found that the photometric-z quasar and galaxy samples expected from the SDSS would perform well as low-z tracers. However, the complete SDSS survey would be required to generate sufficient signal-to-noise. We also find that the ISW effect does not provide enough signal-to-noise to make use of these methods.


\acknowledgements

We are very grateful to Donald Schneider for providing the radial distribution of quasars with $i'<19$ from his SDSS simulation; Alexandre Refregier for useful discussions on the SZ effect, cross-correlating FIRST ellipticities, and the bias of the intracluster gas; Xiaohui Fan and Gordon Richards for information on the SDSS quasar sample; Michael Strauss, Daniel Eisenstein, Andrew Connolly and Rita Kim for information on the SDSS galaxy sample; and Mark Jackson for proofreading this manuscript. HP and DNS were partially supported by the MAP/MIDEX program and NASA NAG5-7154.

\clearpage

\begin{deluxetable}{lccc}
\tablecolumns{4}
\tablecaption{$(S/N)^2$ from the Cross-Correlations. \label{tab1}}
\tablewidth{0pt}
\tablehead{
\colhead{$(S/N)^2$ $(h=0.7, \Omega_0=0.35, \Lambda = 0.65, T_e^0 = 0.5$ keV$)$} 
& \colhead{SZ Effect} & \colhead{Weak Lensing} & \colhead{ISW Effect}
}
\startdata
Photometric galaxies ($r'<21, b_g=1$)  	& 140 & 29 & 6 \nl
All galaxies ($r'<23, b_g = 1$)         & 175 & 77 & 10 \nl
Spectrometric quasars ($i'<19, b_q=3$) 	& 4 & 17 & 3 \nl
Photometric quasars ($i'<21, b_q=3$)   	& 31 & 125 & 8 \nl
FIRST objects ($b_f=2$)          	& $<1$ & $<1$ & $<1$ \nl 
2MASS galaxies\tablenotemark{a}		& 108 & 3 & 1 \nl	
\cutinhead{$(S/N)^2$ $(h=0.76, \Omega_0=0.3, \Lambda = 0.7, T_e^0 = 0.5$ keV$)$}
Photometric galaxies ($r'<21, b_g=1$)	& 124 & 25 & 7  \nl
All galaxies ($r'<23, b_g=1$)		& 158 & 67 & 13 \nl
Spectrometric quasars ($i'<19, b_q=3$) 	& 4 & 15 & 3 \nl
Photometric quasars ($i'<21, b_q=3$)   	& 31 & 119 & 11 \nl
FIRST objects ($b_f=2$)                 & $<1$ & $<1$ & $<1$ \nl 
2MASS galaxies\tablenotemark{a}		& 93 & 2 & 1 \nl	
\enddata
\tablenotetext{a}{Estimated signal-to-noise values for the 2MASS survey (\cite{che98}) are given as a comparison. This all-sky survey will contain $<10^6$ galaxies brighter than $K_s \sim 13.5$ (corresponding roughly to $r' \sim 17$), and about three quarters of the sky will be useful for the purpose of cross-correlating with the CMB. The complete 2MASS dataset will be available much sooner than that of the SDSS.}
\end{deluxetable}



\clearpage


\clearpage


\clearpage


\end{document}